\documentclass[prl,twocolumn,amsmath,amssymb,floatfix,footinbib,bibnotes]{revtex4-1}
\pdfoutput=1

\usepackage[colorlinks=true,citecolor=blue,linkcolor=blue]{hyperref}
\usepackage{amsmath}
\usepackage{graphicx}
\usepackage{dsfont}
\usepackage{changepage}
\usepackage{fancyhdr}
\usepackage{amsthm, amssymb}
\usepackage{floatflt}
\usepackage{float}
\usepackage{array}
\usepackage[usenames,dvipsnames]{color}

\newcommand{\be}{\begin{align}}
\newcommand{\ee}{\end{align}}

\def \ua{{\uparrow}}
\def \da{{\downarrow}}
\def \be{\begin{equation}}
\def \ee{\end{equation}}
\def \ba{\begin{array}}
\def \ea{\end{array}}
\def \bea{\begin{eqnarray}}
\def \eea{\end{eqnarray}}
\def \nn{\nonumber}
\def \L{{\Lambda}}
\def \a{{\alpha}}
\def \t{{\theta}}
\def \b{{\beta}}

\def \D{{\Delta}}
\def \d{{\delta}}
\def \w{{\omega}}
\def \s{{\sigma}}
\def \f{{\varphi}}

\def \ve{{\varepsilon}}

\def \G{{\Gamma}}
\def \z{{\zeta}}

\def \ba{\begin{align*}}
\def \ea{\end{align*}}

\newcounter{indice}

\def \mrm{\mathrm}

\def \bs{\boldsymbol}
\def \mc{\mathcal}

\begin{document}

\title{Odd-parity superconductivity near an inversion breaking quantum critical point \\ in one dimension }

\author{Jonathan Ruhman, Vladyslav Kozii and Liang Fu \\
{\small \em Department of Physics, Massachusetts Institute of Technology, Cambridge, MA 02139 USA}}
\begin{abstract}
We study how an inversion-breaking quantum critical point affects the ground state of a one-dimensional electronic liquid with repulsive interaction and spin-orbit coupling. We find that regardless of the interaction strength, the critical fluctuations always lead to a gap in the electronic spin-sector.
The origin of the gap is a two-particle backscattering process, which becomes relevant due to renormalization of the Luttinger parameter near the critical point.
The resulting spin-gapped state is topological and can be considered as a one-dimensional version of spin-triplet superconductor.
Interestingly, in the case of a ferromagnetic critical point the Luttinger parameter is renormalized in the opposite manner, such that the system remains non-superconducting.
\end{abstract}
\maketitle
\noindent
{\it Introduction --}
Enhancement of superconductivity in the vicinity of a quantum critical point (QCP) has been studied extensively in strongly correlated materials~\cite{Scalapino2012,sachdev2010quantum,Taillefer2010,Shibauchi2014}. Theoretically, this phenomenon was studied for the case of a nematic~\cite{Metlitski2015,Lederer2015,dumitrescu2015superconductivity,schattner2016,li2016makes}, charge ordering~\cite{Chubukov2015} and antiferromagnetic QCP~\cite{berg2012sign,li2015nature,schattner2015competing}.
Recently, it was proposed that the quantum fluctuations associated with an inversion-breaking QCP will also lead to superconductivity in systems with strong spin-orbit coupling~\cite{Kozii2015,Wang2016}. The simplest example would be a paraelectric metal undergoing a quantum transition into an electrically polarized state (i.e. a "ferroelectric metal"~\cite{Anderson1965,shi2013ferroelectric}).
This kind of transition has been observed in metallic compounds such as SrTiO$_3$~\cite{deLima2015} and Cd$_2$Re$_2$O$_7$~\cite{Castellan2002,yamaura2002low}, and was theoretically associated with superconductivity~\cite{Kozii2015,Edge2015,Wang2016}.
The question is how the inversion-breaking QCP affects the electronic states in these materials.

Fluctuations near an inversion-breaking QCP have been proposed to give rise to odd-parity superconductivity~\cite{Kozii2015,Wang2016}. An odd-parity superconductor is formed when the pair wavefunction is odd under inversion, which typically leads to topological superconductivity~\cite{fu2010odd,Sato2010}, for example the B-phase of He-3~\cite{Legget1975}.

Evidence for odd-parity superconductivity were found in doped Bi$_2$Se$_3$~\cite{hor2010superconductivity,Kriener2011,liu2015superconductivity,qiu2015time}, via NMR~\cite{matano2016spin} and specific heat~\cite{yonezawa2016thermodynamic} measurements, which agree with theoretical proposals~\cite{fu2010odd,Nagai2012,fu2014odd}.
The superconducting state was shown to be correlated with structural transitions induced by external pressure~\cite{kirshenbaum2013pressure}, which may also be induced internally by the doping process~\cite{Ribak2016}.
Thus, it is interesting to understand the interplay between structural transitions and superconductivity in topological materials.

In this paper we study the impact of an inversion breaking structural transition on metallic states in one-dimension.
The advantage of one dimension is that we have a good description of the interacting electronic state in terms of a Luttinger liquid (LL).
It is important to note that one-dimensional superconductivity differs from higher dimensional superconductivity. Fluctuations prevent the establishment of true long-range order; instead, the order parameter exhibits only power law correlations, while spin excitations become gapped. This state is known as a {\it Luther-Emery liquid} (LEL)~\cite{Luther1974}.

For concreteness we study a specific model. The model consists of two electronic wires coupled to a soft transverse optical phonon that undergoes a transition into a polarized state, where it breaks inversion between the wires.
We note, however, that the low-energy theory we obtain is generic and describes any spin-orbit coupled system, which is coupled to an inversion breaking transition in one dimension.
We tune the phonon through an inversion-breaking QCP and find that despite the repulsive electron-electron interactions there is always a region, close enough to the critical point, where a LEL is formed.
The electron-electron interactions are crucial for the formation of the gap; they provide a backscattering term, which becomes relevant near the critical point.
We also show that the most divergent superconducting order parameter is odd under inversion, and therefore we identify this state as the one-dimensional version of an odd-parity superconductor. Interestingly, this type of LEL was shown to be a {\it gapless topological state} protected by time-reversal symmetry~\cite{Keselman2015}, and similar to gapped states~\cite{Keselman2013,Haim2014,Haim2016,Klinovaja2014}.
Note that an analogues construction with a ferromagnetic QCP, where time-reversal symmetry is broken instead of inversion, does not lead to superconductivity.

\begin{figure}[t]
 \begin{center}
    \includegraphics[width=1\linewidth]{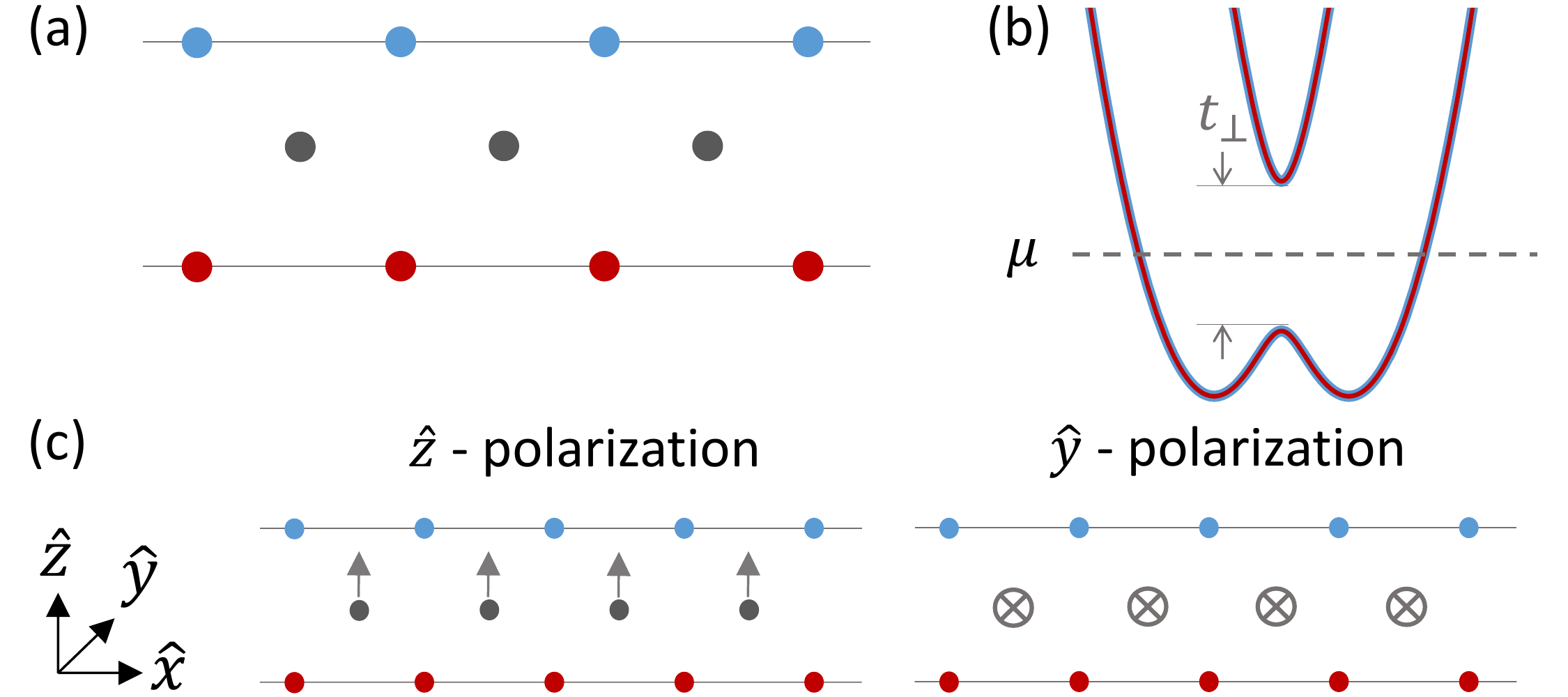}
 \end{center}
\caption{(a) The microscopic model: Two fermionic chains (blue and red) with strong Rashba spin-orbit coupling resulting from the electric field induced by the ions between the wires (grey). (b) The two helical bands $\nu = \pm$. The Fermi energy $\mu$ is taken to be in the gap which is opened by the inter-wire tunneling $t_\perp$, and therefore there are only two helical modes at the Fermi level. (c) The two inversion-breaking distortions of the ions between the wires.  }
 \label{setup:fig}
\end{figure}

{\it Model -- }
We consider a minimal model for a one-dimensional metal with inversion symmetry and spin-orbit coupling. The model consists of two wires, which are interchanged under inversion and thus have Rashba spin-orbit coupling of opposite signs, as shown in Fig.~\ref{setup:fig}(a). The structural inversion-breaking transition is considered to be due to an ionic distortion between these wires, represented by the grey sites in Fig.~\ref{setup:fig}(a). The electrons are coupled to the ions through the local deformation potential generated by the distortion.

We now elaborate on each ingredient of the model starting with the two fermionic wires, which are described by the Lagrangian
\begin{align}
\mc L_f = &\sum_{j=1,2} \psi_{j}^\dag \left[\partial_\tau-{\partial_x^2 \over 2m}\pm\a(-i \partial_x) \s^y  - \mu \right] \psi_j  \nn\\
-&t_\perp \left(\psi_1^\dag \psi_2 +\psi_2^\dag \psi_1\right) + V_{\text{int}}\,,\label{Lf}
\end{align}
where $\psi_1 = (\psi_{1\ua},\psi_{1\da })^T$ and $\psi_2 = (\psi_{2\ua },\psi_{2\da })^T$ are the fermionic fields of the two wires, $m$ is the band mass, $\a$ is the strength of Rashba spin-orbit coupling and $\pm$ correspond to $j=1,2$, respectively. $\mu$ is the chemical potential. The term $V_{\text{int}}$ describes a generic $S_z$ conserving local repulsive interaction, and $\s^i $ are the Pauli matrices in the spin space.

Tunneling between the wires, $t_\perp$, opens a gap at the $\G$-point, as shown in  Fig.~\ref{setup:fig}(b). We consider the case $|\mu|<t_\perp$, where only two points cross the Fermi energy. These two modes are denoted by their helicity $\nu=\pm$, which is interchanged under inversion.

Next, we consider the phonon mode that becomes soft at the QCP.
This mode describes the motion of the ions located in between the two fermionic wires, as shown in Fig.~\ref{setup:fig}. These ions are localized, but may fluctuate around their equilibrium positions. We consider tuning these ions to a critical point where they become soft and condense in a different configuration, which breaks the inversion symmetry between the wires~\cite{Anderson1965}.

Among the three (one longitudinal and two transverse) phonon modes, only the transverse modes become soft at the transition~\cite{Roussev2003}.
The transverse motion is decomposed into two inversion-breaking polarizations, $\hat y$ and $\hat z$ [see Fig.~\ref{setup:fig}(c)]. The electronic wires, lying in the $ x z$-plane, lift the degeneracy between these modes, such that one can focus on the lower energy one, which becomes soft near the QCP. This mode can be described by a scalar field
\be
\mc L_b = {1 \over 2}\f\,  \mc G_b^{-1}\,\f\,,\label{Lb}
\ee
where $\mc G_b = \left[-  \partial_\tau^2 -  \partial_x^2 + {r}\right]^{-1}$, $\f$ describes the ionic displacement, $r$ is the mass term and the velocity has been set to unity for convenience. In what follows we consider $r$ as a tuning parameter which describes softening of an optical phonon frequency.
%
%Classically, the Ising theory given by Eq.~(\ref{Lb}) undergoes a transition at $r = 0$. The interaction term $U$ generates quantum fluctuations, which shift the transition point to a negative value $r \approx - (3U/\pi)|\ln(U/U^*)|$, where $U^*$ is a non-universal parameter (see Ref.~\cite{Podolsky2014}). However, as we show, the coupling between the fermionic and Ising sectors also shifts the transition point. In the limit of small $U$ the latter effect is much stronger, shifting the transition point to $r = r_c>0$. This allows us to neglect $U$ completely hereafter.

We now turn to the coupling between the phonon mode and fermions. This coupling originates from the deformation potential induced by the motion of the ions between the  fermionic wires, as shown in Fig.~\ref{setup:fig}(c). For $\hat z$-polarization, the electrostatic field generates a chemical potential difference between the wires, leading to the coupling
\be
\mc L_{fb} = -\lambda\, \f\, \left(\psi_1^\dag  \psi_1 -\psi_2^\dag  \psi_2\right). \label{Lbf}
\ee
In what follows we consider only this polarization. We note that in the case of the $\hat y$-polarization, shown on the right panel of Fig.~\ref{setup:fig}(c), the lattice distortion generates a Rashba-like effect for tunnelling between the wires, leading to a coupling of the form $\mc L'_{fb} = -i\lambda\, \f\, \left(\psi_1 ^\dag \s^x \psi_2 - \psi_2 ^\dag \s^x \psi_1\right)$. This expression can be reduced to Eq.~(\ref{Lbf}) by a unitary transformation, so the case of $\hat y$-polarization leads to the same results.

{\it Analysis of the model near the QCP --}
To analyze the model given by Eqs.~(\ref{Lf})-(\ref{Lbf}) we bosonize the two fermionic modes crossing the Fermi energy $\psi_{\ve,\nu} \simeq (F_\nu / \sqrt{2 \pi a}) \exp(i \ve k_F^{\nu} x)  \exp[-i\left( \ve \t_\nu -\phi_\nu \right)]$. Here $F_{\nu}$ are Klein factors, $a  $ is the short distance cutoff, $\ve = R,L$ and $\nu=\pm$ denote the chirality and helicity of the modes, respectively ($R,L$ correspond to $\ve = +, -$, respectively). Bosonic fields $\t_\nu$ and $\phi_{\nu'}$ obey commutation relations $\left[ \t_\nu(x), \phi_{\nu'}(x') \right] = i \pi \delta_{\nu \nu'}\,\Theta(x'-x)$, where $\Theta(x)$ is the Heaviside step function. In these notations, $\rho_\nu = -(1/ \pi)\partial_x \t_\nu$ and $J_\nu = (1/ \pi)\partial_x \phi_\nu$ are the charge and current densities of the helical band $\nu$, respectively (the uniform part of density is omitted). Because of the helical structure of the bands, spin density $\rho_{\nu}^s$ and spin current density $J_{\nu}^s$ can be expressed through charge components as $\rho_{\nu}^s = -\nu J_{\nu}$ and $J_{\nu}^s = \nu \rho_\nu$.

After the bosonization, the fermionic Lagrangian (\ref{Lf}) becomes a sum of two decoupled LLs describing charge and spin degrees of freedom:
\begin{align}
\mc L_f =\mc L_\rho +\mc L_\s = {1\over 2}\sum_{\eta = \rho ,\s}
\bs \Phi_\eta^T
\,\mc G_\eta ^{-1}\,
\bs\Phi_\eta + {g\over 2 \left(\pi a\right)^2} \cos \sqrt{8} \t_\s , \label{Lf:bos}
\end{align}
where
$\mc G_\eta^{-1} = {1\over \pi}\left(\begin{matrix}
-u_\eta K_\eta \partial_x^2  &  {i}\partial_\tau \partial_x   \\
 {i}\partial_\tau \partial_x & -u_\eta K_\eta^{-1} \partial_x^2
\end{matrix}\right) $
is the bare Green's function and $\bs \Phi_\eta = (\phi_\eta , \t_\eta)^T$. Here we defined 'charge' and 'spin' variables
 $\t_{\rho} \equiv (1/\sqrt{2})\left(\t_+ + \t_-\right)$,  $\phi_{\rho} \equiv(1/\sqrt{2}) (\phi_+ + \phi_-)$,  $\t_{\s} \equiv  (1/\sqrt{2})\left(\phi_+ - \phi_-\right)$ and $\phi_{\s} \equiv(1/\sqrt{2})(\t_- - \t_+)$ that obey the same commutation relations as $\t_\nu$ and $\phi_{\nu'}$. Coefficients $K_\rho$ and $K_\s$ are Luttinger parameters for the charge and spin sectors, respectively. We consider repulsive fermionic interaction, which generically corresponds to $K_\rho<1$ and $K_\s>1$~\cite{giamarchi2004quantum}.
 Coefficients $u_\rho$ and $u_\s$ are velocities of charge and spin excitations. Expressions for $u_\eta$ and $K_\eta$ in terms of microscopic parameters of Eq.~\ref{Lf} can be found in Ref.~\cite{Orignac1997}.

The cosine in Eq.~(\ref{Lf:bos}) originates from the BCS scattering processes, , i.e., from the term $\psi_{R,+} ^\dag \psi_{L,+} ^\dag \psi_{L,-}\psi_{R,-} + \mrm{H.c.}$, which describes pair hopping between the fermionic wires.
For small bare $g$, the behavior of this term in the renormalization group (RG) sense is determined by the Luttinger parameter $K_\s$. In the case of repulsive interactions, $K_\s>1$, cosine term is irrelevant and flows to zero. Attractive interaction (corresponding to $K_\s<1$), on the contrary, guarantees that $g$ flows to the strong coupling regime, and pins the variable $\t_\s$ to one of the minima of the cosine. This state is known as the  LEL~\cite{Luther1974} and is characterized by a {\it spin gap}.

We consider repulsive electron-electron interaction, so one could naively expect that the cosine term becomes irrelevant. However, as we show, close enough to the critical point the cosine  {\it always} becomes relevant rendering the spin sector gapped, regardless of the strength of the repulsive interaction. To demonstrate it explicitly, we rewrite the coupling term (\ref{Lbf}) in the bosonized form:
\be
\mc L_{fb} =-{\sqrt 2\,\lambda\over \pi }\, \f\, \partial_x \phi_\s \,.\label{Lbf:bos}
\ee
As expected, the phonon field couples to the total spin current density $J_\s = (\sqrt 2/ \pi)\partial_x \phi_\s $. It is important to note that the coupling to spin current, Eq.~(\ref{Lbf:bos}), vanishes in the absence of spin-orbit coupling~\cite{Kozii2015}. In this case the gaplesness of the spin-sector of the fermions is protected by SU(2) symmetry~\cite{giamarchi2004quantum}.
%thus, indeed, breaking inversion symmetry in the ordered phase.

Now all ingredients are ready for constructing the effective theory that elucidates the main results of this work. First we integrate out the massive phonon mode Eq.~(\ref{Lb}). Formally, this procedure is equivalent to the shifting $\tilde \f = \f - (\sqrt{2}/ \pi) \mc G_b\partial_x \phi_\s$, which demonstrates that the phonon field is {\it locked} to the spin current which is odd under inversion. As a result, the Lagrangian of the spin sector in Eq.~(\ref{Lf:bos}) becomes
\begin{align}
\mc L_{\s} \rightarrow \mc L_{\s}   -{ \lambda^2 \over \pi^2}\,\partial_x \phi_\s\,\mc G_b\, \partial_x\phi_\s   \label{Lf_2}
%\\
%&\approx {1\over 2}\bs \Phi_\s^T  \mc G_\s ^{-1} \bs \Phi_\s  - { \lambda^2\over \pi^2 r}\, \left( \partial_x \phi_\s \right)^2+{g\over 2 \left(\pi a\right)^2} \cos \sqrt{8} \t_\s \nn .
\end{align}
Note that because the real ordering transition occurs at a finite value of the tuning parameter $r$, the bosonic field $\varphi$ is massive and therefore can be integrated out without generating non-analyticities.

At long wavelength and low energy, the resulting effective theory (\ref{Lf_2}) has the same form as Eq.~(\ref{Lf:bos}) with a renormalized spin stiffness, which is suppressed by the critical phonon fluctuations. Consequently, the Luttinger parameter $K_{\sigma}$ and spin velocity $u_\s$ also become significantly suppressed:
\be
\tilde K_\s(r) = K_\s \sqrt{1-{r_c\over  r}}\;\; , \;\; \tilde u_\s(r) =u_\s \sqrt{1-{r_c \over  r}}\, , \label{ren_stiff}
\ee
where $r_c \equiv {2\lambda^2 /\pi u_\s K_\s}$.

We identify two critical values of mass term $r$:

\noindent
(i) First, for values of $ r$ smaller than $r_* \equiv r_c / (1-K_\s^{-2}) $ the renormalized Luttinger parameter $\tilde K_\s(r)$ becomes smaller than 1. Therefore, $r_*$ marks the Berezinskii-Kosterlitz-Thouless (BKT) transition point between the gapped LEL and the gapless LL. We note that $r_*$ is the critical value for a vanishing $g$. The case of finite $g$ appears in the SI~\cite{SI}.

We emphasize that, as opposed to the conventional scenario where the LEL is realized in systems with attraction between electrons (implying $K_{\sigma}<1$, $g<0$), in our theory it naturally appears in the presence of repulsive interaction, i.e. $g>0$. This type of spin-gapped liquid was studied in detail in Refs.~\cite{giamarchi1986theory,giamarchi1988theory}. More recently, it was pointed out in Ref.~\cite{Keselman2015} that a LEL with $g>0$ is a {\it gapless symmetry protected topological state}. In this case, the most divergent superconducting correlations are of spin triplet type. To see this, one may compare the correlations of the triplet order parameter, $ O_{t} = \psi_{R,+}^\dag \psi_{L,+}^\dag-\psi_{L,-}^\dag \psi_{R,-}^\dag \propto  e^{-i\sqrt 2\phi_\rho}\sin {\sqrt{2}\t_\s } $, with those of the singlet, $O_{s} = \psi_{R,+}^\dag \psi_{L,+}^\dag+\psi_{L,-}^\dag \psi_{R,-}^\dag \propto  e^{-i\sqrt 2\phi_\rho}\cos {\sqrt{2}\t_\s} $. For $g>0$, the field $\t_\sigma$ is pinned to $(\pi/\sqrt{8}) + (\pi n/\sqrt{2}) $ ($n$ is integer), such that the correlations of $\mc O_s$ are exponentially suppressed while correlations of $\mc O_t$ exhibit quasi-long-range order (note that here the operators are written in the helical basis where spin is locked to momentum). It should be noted, however, that in this case the most divergent order parameter is the spin-density wave. This is due to the correlations in the charge sector which are controlled by $K_\rho<1$ (repulsive interactions)~\cite{giamarchi1986theory,giamarchi1988theory}. However, in the last part of this paper we couple an array of wires and show that the two-dimensional version of our model can realize odd-parity superconductivity.

(ii) Next, if the tuning parameter $r$ is further decreased, the system undergoes the second phase transition. Indeed, at $r \leq r_c$, the stiffness of the field $\phi_\s$ becomes negative, meaning that  $r_c$ marks the {\it classical} transition point to the inversion-breaking phase where the system develops a spontaneous spin current. To stabilize the theory, as usual, the higher order term, $ \left(\partial_x \phi_\s \right)^4$, that has been neglected in the derivation of Eq.~(\ref{Lf:bos}) must be taken into account. This term may have different microscopic origin\footnote{  This term originates from higher order interaction terms. For similar discussion see~\cite{yang2004ferromagnetic}.}.

To understand the nature of the transition at $r = r_c$ we assume that in the vicinity of $r_c$ the system is already deep in the spin-gapped state, so that the cosine term in Eq.~(\ref{Lf_2}) can be expanded near one of its minima, leading to an effective gap $\D$. We will justify this assumption later, by explicitly calculating $\D$ using the variational principle.
In this case we can integrate out the field $\t_\s$ in Eq.~(\ref{Lf_2}) and express the resulting Lagrangian in terms of the spin current, $J_\s = (\sqrt{2}/ \pi)\partial_x \phi_\s$. In the long wavelength limit (which implies $r\gg  q, \omega$ and $\D \gg u_\s q$), it becomes.
\be
\mc L_{\mrm{Ising}} = {1\over 2}\,J_\s \left(-{\partial_\tau^2 \over \z} -  \rho_\s\partial_x^2+\a \right) J_\s +  V J_\s^4\,, \label{Ising}
\ee
where $\zeta = 2\D^2 r^2/ \pi u_\s K_\s (r^2+\D^2 r_c)$, $\rho_\s = {\pi} u_\s K_\s r_c /2r^2$ and $\a = {\pi} u_\s K_\s (1- r_c/r)/2$.
The Lagrangian~(\ref{Ising}) describes the Ising transition in 1+1 dimensions. Here $\a$ plays the role of the tuning parameter.
The term $ V$ is responsible for quantum fluctuations, which shift the transition point.
We note that in the absence of particle-hole symmetry the Ising theory, Eq.~(\ref{Ising}), is also coupled to the charge sector by $\mc L_{\rho \s} = {\lambda_\rho\over \pi} \partial_x \t \, J_s^2$, which modifies the nature of the transition~\cite{Sitte2009,Alberton2016}. However, it couples the gapless-charge to gapped-Ising modes and thus does not shift the transition point or change the nature of the Ising phases.

\begin{figure}[t]
 \begin{center}
    \includegraphics[width=1\linewidth]{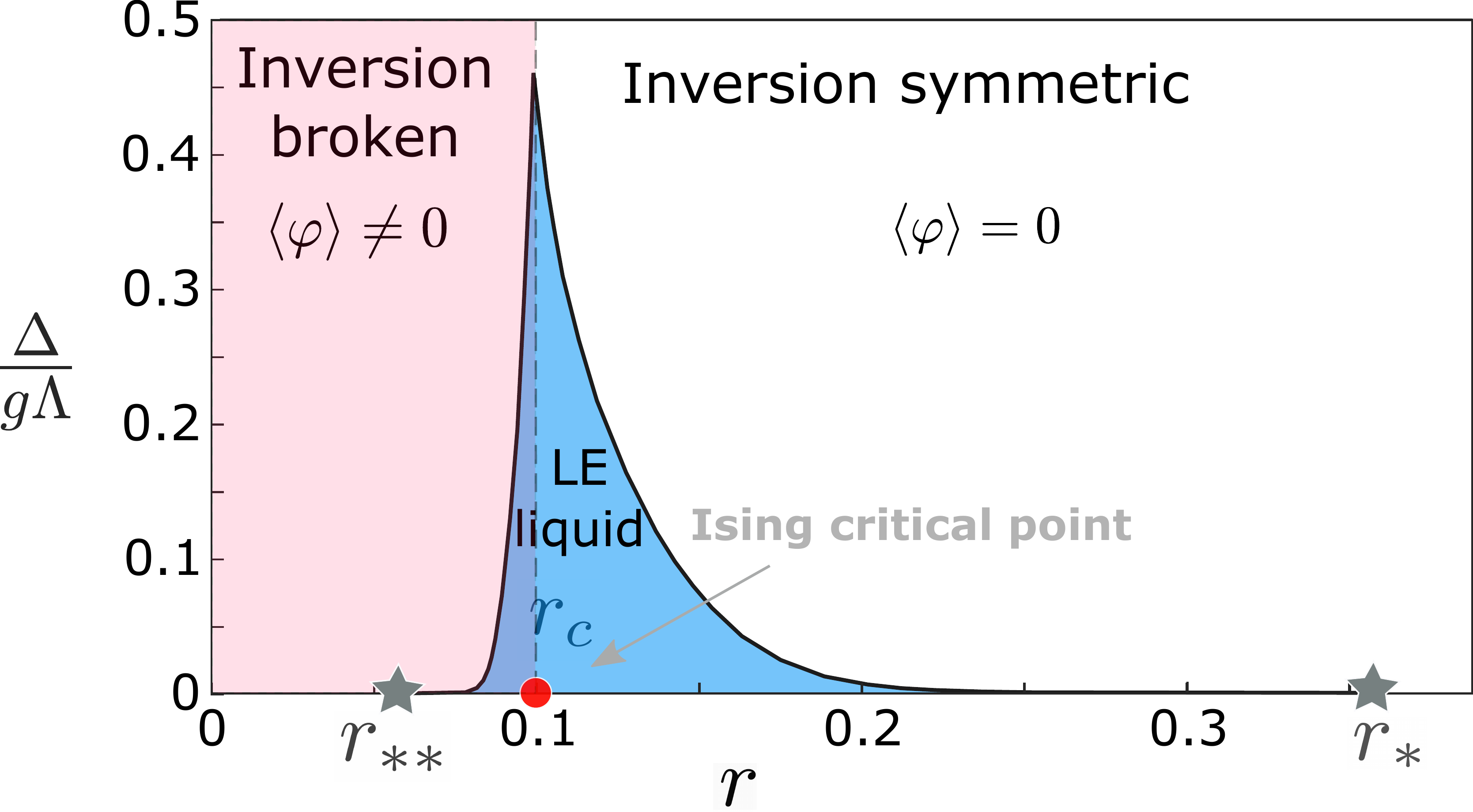}
 \end{center}
\caption{Fermionic gap $\D$ obtained from the variational calculation vs. the tuning parameter $r$ for $u_\s = 3/2$, $g = 0.2$, $r_c = 0.1 \L^2$, $K_\s = 1.2$ (repulsive interactions) and $\L$ is the ultra-violet cutoff of the theory. The two stars at $r = r_*$ and $r = r_{**}$ denote the BKT transition points between the topological LEL and the gapless LL state. $r_c$ denotes the Ising critical point.  }
 \label{gap:fig}
\end{figure}

We now contrast the results we have obtained for inversion-breaking transition with the ferromagnetic case considered in Ref.~\cite{yang2004ferromagnetic}. In the latter, the ferromagnetic fluctuations couple to the spin density as opposed to spin current, implying that  $\partial_x \t_\s$ substitutes $\partial_x \phi_\s$ in Eq.~(\ref{Lbf:bos}). The renormalized Luttinger parameter then equals $\tilde K_\s(r) = {K_\s/\sqrt{1-(r_c K_\s^2/ r)}}$, which is only enhanced in the vicinity of the QCP, making the cosine term even more {\it irrelevant}. Therefore, in the case of a ferromagnetic transition, the fermionic spin sector remains gapless all the way to the QCP, which is characterized by the dynamical exponent $z=2$, as opposed to the Ising transition in our case. In this case the superconducting correlations are not divergent.

{\it Calculation of the spin gap --}
We now turn to explicitly calculate the magnitude of the spin gap, $\D$, as a function of the tuning parameter $r$. For this purpose we employ the variational principle, which is known to capture the qualitative behavior of the gapped LEL~\cite{giamarchi2004quantum}. We introduce the variational action $\mc L_{\s}^V =  {1\over 2}\bs \Phi_\s^T  \mc G_\s ^{-1} \bs \Phi_\s  -{ \lambda^2 \over \pi^2}\,\partial_x \phi_\s\,\mc G_b\, \partial_x\phi_\s  +{\D^2 } \t_\s^2/u_\s K_\s  $, where $\D$ is the variational parameter, which represents the gap of spin excitations [the same $\D$ appears in Eq.~(\ref{Ising})].
Therefore, we minimize the free energy corresponding to Eq.~(\ref{Lf_2}) with respect to the $\D$~\cite{SI,giamarchi2004quantum}.
On the ordered side of the transition ($r < r_c$) the spin stiffness is negative. In this case we expand around the broken symmetry state, taking into account the higher order term presented in Eq.~(\ref{Ising}).
Consequently, another BKT transition occurs on the ordered side, at $r = r_{**} \equiv r_c /\left[1+\left(2K_\s^2\right)^{-1} \right]$~ \cite{SI}.

We note that the variational approach breaks down at two points: near $r = r_*$ and near $r  = r_c$.  In the former, the gap vanishes like $\D \approx \sqrt{r_*}  \,\exp\left({- {\pi \over A \sqrt{1-r/r_*}}}\right)$, where $A$ is defined in the SI~\cite{SI}. In the latter, the quartic term $V$ in Eq.~(\ref{Ising}), which we have neglected in this calculation, becomes important in the regime $|r-r_c| \sim \rho_\s V/\z$~\cite{Podolsky2014}.

The results of the variational calculation~\cite{SI} are summarized in Fig. \ref{gap:fig}, where we plot the gap, $\D$, as a function of the tuning parameter $r$. Starting from the disordered side and reducing the tuning parameter $r$ towards the critical value $r_c$, the fermionic sector first undergoes a BKT transition into the LEL at $r = r_*$. Then, at $r = r_c$ the Ising transition occurs and inversion becomes spontaneously broken, where the gap reaches its maximal value. Finally, at $r =r_{**}$, there is another BKT transition to the gapless LL state.

{\it 2D wire construction --}
We now extend our results to two dimensions by discussing the phase diagram of an array of wires, which are individually described by Eqs.~(\ref{Lf_2}) and the gapless charge sector. Close to the QCP and in the weak coupling limit, $\t_\s$ is pinned on each wire and there is a finite gap $\D$.
%the strength of the coupling between the wires is weak enough such that Eq.~(\ref{Lf_2}) is a good starting point to describe the two-dimensional system. We also assume that we are close enough to the QCP such that $\t_\s$ is locked to $\pm \pi/\sqrt{8}$.
In this case, for weakly coupled wires we are left with an array of LL in the charge sector, $\mc L_j = {1\over 2} \left(\bs \Phi_\rho^j\right)^T \mc G_\rho^{-1}\bs \Phi_\rho^j$, interacting by
\begin{align}
\mc L_{j,j+1} =& {1 \over 2\pi }   \left(W_\phi \,\partial_x \phi_\rho^j \partial_x \phi_\rho^{j+1} + W_\t \,\partial_x \t_\rho^j \partial_x \t_\rho^{j+1}\right)\nn
\\&+ g_\phi\cos\sqrt{2} \left(\phi_\rho^j - \phi_\rho^{j+1} \right)+ g_\t\cos \sqrt 2 \left(\t_\rho^j - \t_\rho^{j+1}  \right)\,.\nn
\end{align}
$W_\phi$ and $W_\t$ describe forward scattering (for repulsive interaction $W_\phi < W_\t$). The cosine term $g_\phi$ is proportional to the strength of pair hopping between the wires and $g_\t$ results form  wire-number conserving processes.

\begin{figure}[h]
 \begin{center}
    \includegraphics[width=0.8\linewidth]{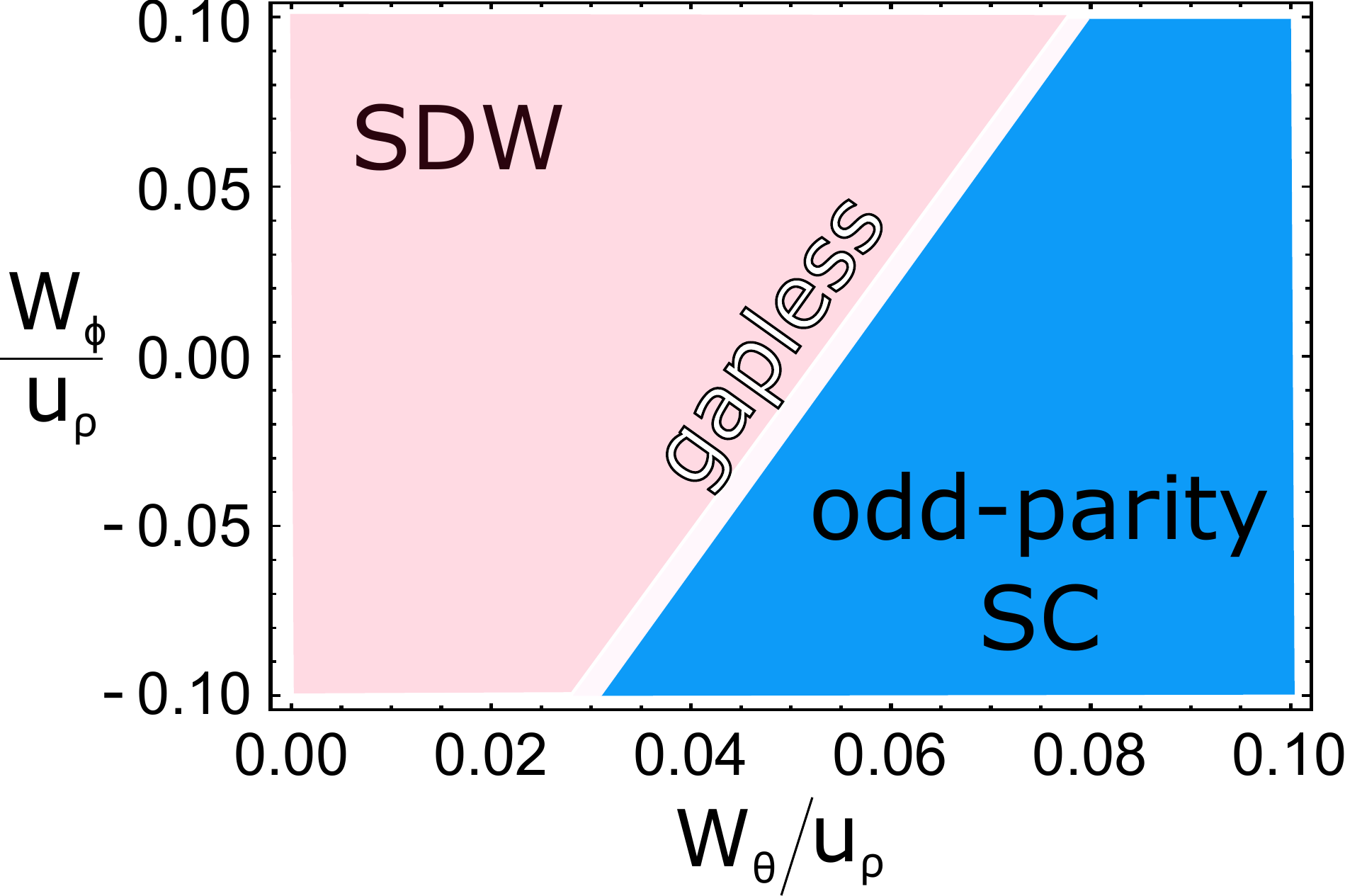}
 \end{center}
\caption{Phase diagram obtained from the RG equations in Refs.~\cite{Emery2000,Vishwanath2001}. The blue region denotes odd-parity superconductivity, white slither denotes gapless sliding LL state and pink denotes a spin ordered state.}
 \label{phase_diagram:fig}
\end{figure}

This model was studied in Refs.~\cite{Emery2000,Vishwanath2001,Mukhopadhyay2001}. It was shown that a superconducting phase, where $g_\phi$ is the most relevant perturbation, exists over a wide range of parameters in the case of spin-gapped wires~\cite{Mukhopadhyay2001}.
In Fig.~\ref{phase_diagram:fig} we plot the phase diagram based on the scaling equations for $g_\phi$ and $g_\t$~\cite{Mukhopadhyay2001}. However, in our case the spin variable $\phi_\s$ is locked to $\pm \pi/\sqrt{8}$ rather than $0,\pi$. As a result the two gapped phases are a spin-density wave and odd-parity superconductor rather than a charge density wave and even-parity superconductor. The novel aspect of this result is that inversion breaking clearly enhances the odd-parity channel over the even-parity one.

{\it Conclusions --}
We have shown that a one-dimensional electronic liquid with strong spin-orbit coupling and repulsive interactions coupled to an inversion breaking QCP develops a spin gap close enough to the transition due to the establishment of a paired state.
We have argued that this state indicates the emergence of odd-parity superconductivity near inversion breaking QCP.
We have also shown that inversion breaking is distinct from the ferromagnetic transition, where the spin sector remains gapless.

{\it Acknowledgments -- }
We are grateful to Eduardo Fradkin, Leo Radzihovsky, Emanuele Dalla Torre and Samuel Lederer for insightful discussions.
This project was funded by the DOE Office of Basic Energy Sciences, Division of Materials Sciences and Engineering under Award de-sc0010526 (LF and VK),
JR acknowledges a fellowship from the Gordon and Betty Moore Foundation under the EPiQS initiative (grant no. GBMF4303).

\onecolumngrid

{\center \Large \bf Supplementary material for "Odd-parity superconductivity near an inversion breaking quantum critical point  in one dimension"}
\newline

This supplementary material consists of two sections. In Section I, we elaborate on the variational method used to calculate the gap $\D$. This method breaks down near the BKT transition where $\D $ vanishes. Therefore in Section II we use the RG flow equations for the BKT transition to calculate $\D$ near the transition point $r = r_*$.

\section{Details of the variational calculation of the spin gap}
In this section we elaborate on the variational method we used to calculate the gap in Fig.~2 of the main text. The calculation follows Ref.~\cite{giamarchi2004quantum}.
We start by introducing the variational Lagrangian $\mc L_{\s}^V = (1/2)\bs \Phi^T_\s \mc G_{V}^{-1} \bs \Phi_\s$, where
\be
\mc G_V^{-1} = {1\over \pi}\left(\begin{matrix}
u_\s K_\s \left[1 - {r_c \over {\w^2+q^2 + r }}\right]q^2 &  {i}\w q  \\
 {i}  \w q & K_\s^{-1}\left[u_\s  q^2 +{\D^2\over u_\s} \right]
\end{matrix}\right)\, \label{GV}.
\ee
We then minimize the expectation value of the free energy corresponding to the Lagrangian given by Eq.~(6) of the main text. The expectation value is taken with respect to the Gaussian Lagrangian $\mc L_\s^V$, and $\D$ is the variational parameter~\cite{giamarchi2004quantum}.
As a result, the following self-consistent equation for $\Delta$ can be found:
\be
{\D^2 \over \pi u_\s K_\s} = {16 g\over \left(2\pi a\right)^2}\,  \exp\left[- {4\over \b L } \sum_{\w,q}\mc G_{V}^{\t\t} (i\w,q)\right]\,, \label{self_consistent}
\ee
where $G_V ^{\t\t}$ is the $\t \t$ component of the Green's function~(\ref{GV}) given by
\[G_V^{\t\t} (\w,q) =    {\pi K_\s u_\s \left( \w^2+B  \right) \over A B + \left( r_c + A+B \right)\w^2+\w^4} \;\;\;; \;\;\;A\equiv u_\s^2 q^2 +\D^2\;\;\;;\;\;\;B\equiv  q^2 +r-r_c  \,,\]
$L$ is the system size and $\b$ is the inverse temperature.

 To obtain Fig.~2 of the main text we solve Eq.~(\ref{self_consistent}) numerically. In the limit of small $\Delta$, it can be resolved analytically (we consider zero-temperature limit):
\be
\frac{\Delta^2}{\pi u_{\s} K_{\s}} = \frac{4g}{\pi^2 a^2}\left({\tilde\L \over \L_0}\right)^{2 K_\s}\left( \frac{\Delta}{ u_\s \tilde \L} \right)^{2K_\s \sqrt{(r-r_c)/r}}, \label{smallgap}
\ee
where $\tilde \Lambda \sim \sqrt{r}$ is an effective energy scale below which the renormalized Luttinger parameters are described by Eq.~(7) of the main text, and $\L_0 \sim 1/a$ is the ultraviolet cutoff.
This result coincides with the RG calculation deep in the massive phase~\cite{giamarchi2004quantum}, and correctly predicts BKT transition point $r_* = r_c/(1-K_\s^{-2}),$ defined as a point where non-zero solution for $\Delta$ appears.

On the ordered side of the transition, $r<r_c$, the $q=0$, $\w = 0$ part of the $\phi \phi$ component of the matrix (\ref{GV}) becomes negative, signaling the instability to the broken symmetry state. As explained in the main text, in this case we must stabilize the theory with higher order terms of the form $V\left(\partial_x \phi_\s \right)^4$, which are irrelevant near the Luttinger liquid fixed point, but are clearly important here.
We make use of the fact that the mass term of a Ginzburg-Landau theory on the ordered side of the transition $M_o$ can be related to the mass on the disordered side $M_d$ by $M_o = -2 M_d$, and importantly is independent of $V$. Thus, to describe the system in the ordered phase we shift the prefactor of $ (\partial_x\phi_s)^2 $ by $3u_\s K_\s \left(1-r_c / r \right)$. Afterwards, the analysis above can be simply repeated for the regime $r < r_c$.
Consequently, another BKT transition occurs on the ordered side, at $r = r_{**} \equiv r_c /\left[1+\left(2K_\s^2\right)^{-1} \right]$.
This second BKT transition must appear, because deep in the broken symmetry phase the $\f$ degrees of freedom freeze. In this limit the only effect of the bosonic degrees of freedom is to break inversion, thus leading to a helical band structure at the Fermi energy equivalent to a single wire with Rashba spin-orbit coupling.  Such a model does not have a spin gap when the interactions are strongly repulsive (see for example Ref.~\cite{Ruhman2015}).

\section{Two step RG calculation of the spin gap near the BKT transition}
In this section we analyze the BKT transition near $r = r_*$. We first obtain the value of $r_*$ for a finite value of $g$. Then we estimate the single particle gap $\D$ close to $r = r_*$. As explained above, the variational method breaks down at the transition because it does not take into account fluctuations which renormalize $K_\s$~\cite{giamarchi2004quantum}. For this purpose we consider the RG flow equations
 \begin{align}
 &{d y(l) \over dl} = \left[2-2 K(l) \right] y(l) \label{y} \\& {d K(l) \over dl} = -{K^2(l) y^2(l) \over 2}\,.\label{K}
 \end{align}
Here $K(l)$ is the running Luttinger parameter of the spin sector, $y(l) = g(l)/\pi u_\s$ is the dimensional coupling constant and $l = \log \left( \L_0 / \L \right)$ is the RG time. The initial values for these parameters are, therefore, $y(0)= {g / \pi u_\s}$ and $K(0) = K_\s$.

The gap can be estimated from the RG equations in the standard way. Basically, we integrate over the RG equations between the initial point $l = 0$ and $l = l_\D$, where we define the scale $\D = \L_0 e^{- l_\D}$ as the scale at which $y(l_\D) \sim 1$. However, in this case we must separate the integration in two regions:
\newline
\noindent
(i) For $\tilde \L <\L< \L_0$, where $\tilde \L\sim \sqrt{r}$, we are integrating out high energy states which are not affected by the soft phonon mode $\f$, which has a gap of size $\tilde \L$.
Therefore $ K(l)$ is not renormalized by the phonons.
If we assume that $K_\s -1 \gg y(0)/2$ then we are far from the seperatrix and therefore the flow is approximately vertical. Therefore, $K(l) = K_\s$ and the flow is described solely by Eq.~(\ref{y}).
Integrating over Eq.~(\ref{y}) in this region yields
\be y_r= {y_0}\left({\L_0 \over \tilde{\L}} \right)^{2-2K_\s}\,. \label{y_r}\ee
\noindent
(ii) In the second step we integrate in range $\D <\L< \tilde{\L}$. Here $K_\s$ and $u_\s$ become renormalized by the soft phonon mode $\f$, according to Eq.~(7) of the main text
\be
K(\tilde \L)= \tilde K_\s (r)=K_\s\sqrt{1 - {r_c \over r}} \;\;\; ,\;\;\;u_\s \rightarrow \tilde u_\s (r)=u_\s\sqrt{1 - {r_c \over r}} \;\;\; \mrm{and \; therefore}\;\;\;{ y_r \rightarrow {y_0\over \sqrt{1-{r_c\over r}}}}\left({\L_0 \over \tilde{\L}} \right)^{2-2K_\s}\,. \label{KK}
\ee
Since we are interested in analyzing the transition point we seek the value of $r_*$ such that these renormalized parameters land {\it on the seperatrix}. In this case the flow strongly modifies $K$ and we must use both equations.
Demanding that $\tilde K_\s(r_*)=1+y_*/2$ (which defines the sepratrix at the scale $\tilde \L$), where $y_* = y_{r_*}$, we obtain from Eq.~(\ref{KK}) the self-consistent equation for the transition point
\be
r_* = {r_c \over 1- K_\s^{-2} \left[ 1+ {y_*\over 2} \right]^2}\,,
\ee
which reduces to $r_* = r_c / \left( 1-K_\s^{-2} \right)$ in the limit of $y = 0$. In the main text (Fig.~2) we have used $g = 0.2$, $u_\s = 3/2$, $r_c = 0.1 \L^2$ and $K_\s = 1.2$. These parameters lead to $r_* \approx 0.361\ldots$

Defining the parameter $y_\parallel (l)\equiv 2 K(l) -2$ and combining Eqs.~(\ref{y},\ref{K}) one obtains
\be
y = \sqrt{y_{\parallel}^2+\d^2 }\,,\label{y_p}
 \ee
 where $\d^2 =A^2\left( {1-{r / r_*}}\right)$ dictates the flow line, where
  \[A = \sqrt{2 \left( a_2 - a_1\right)y_*}\,,\]
  $a_1 = K_\s r_c / r_*\sqrt{1-r_c / r_*}$ and $a_2 = \left\{K_\s - 1 +{1\over 2} \left[ r_c / \left( r_c - r_* \right)\right]\right\} y_*$.

Plugging Eq.~(\ref{y_p}) in the RG equations Eqs.~(\ref{y},\ref{K}) we obtain
\be
{dy_\parallel (l)\over dl} = - \d^2 - y_\parallel^2(l)\,.
\ee
Integrating this equation between $\tilde \L$ and $\D$ one obtains~\cite{giamarchi2004quantum}
\be
\D \approx \tilde \L \,\exp\left({- {\pi \over A \sqrt{1-r/r_*}}}\right)
\ee
Therefore close to the seperatrix the gap vanishes exponentially fast.
This result indicates that all derivatives of the gap $\D(r)$ vanish at $r =r_*$, signaling that the free energy is analytic to all orders, a well known property of the BKT transition~\cite{kardar2007statistical}.


\begin{thebibliography}{55}
\bibitem [{SI()}]{SI}%
  \BibitemOpen
  \href@noop {} {}\bibinfo {note} {See supplimentary material for details on
  the varaitional calcualtion of the gap and the estiamtion of the gap close to
  the BKT trnasition points, which inclueds Refs.~[45, 46]}\BibitemShut
  {NoStop}%
\end{thebibliography}
\end{document}